\documentclass[aps]{revtex4}
\usepackage{polski}
\usepackage{amsmath,amssymb}
\usepackage{color,graphicx}
\usepackage{multirow}
\usepackage{amsmath,amssymb}
\newcommand{\be}{\begin{equation}}
\newcommand{\ee}{\end{equation}}
\begin{document}
\title{ Nonlocal  random motions and the trapping problem\thanks{Presented at the
27th Marian Smoluchowski Symposium  on Statistical Physics, Zakopane, Poland,
22-26 Sept. 2014}}
\author{Piotr Garbaczewski and Mariusz \.{Z}aba} \affiliation{Institute of Physics,
University of Opole, 45-052 Opole, Poland}
\date{\today }
\begin{abstract}
 L\'{e}vy stable (jump-type) processes are examples of intrinsically
nonlocal random motions. This property becomes a serious  obstacle
if one attempts to model conditions under which a   particular
L\'{e}vy process may be subject to physically implementable
manipulations,  whose ultimate goal is to  confine the random motion
in a spatially finite, possibly mesoscopic  trap. We analyze this
issue for an exemplary case of the Cauchy process in a finite
interval.  Qualitatively, our observations extend to general
jump-type processes that are driven by non-gaussian noises,
classified by the  integral part of the L\'{e}vy-Khintchine formula.
For clarity of  arguments we  discuss, as a reference  model,
  the classic  case of the Brownian motion in the  interval.
    \end{abstract}
\maketitle

\section{Motivation:}

In contrast to the locally defined generator $\Delta  = \partial ^2/\partial x^2$  of
the standard  Brownian motion in $R$ , generators of L\'{e}vy stable processes are spatially nonlocal.
In fact, in the symbolic    notation (a fractional power  $(-\Delta )^{\mu /2}$ is  here replaced by
$|\Delta |^{\mu /2}$) we have
the $\mu \in (0,2)$-stable  generator defined as follows:
\begin{equation}
- |\Delta |^{\mu /2} f(x)  =   \int [f(x+y) - f(x)]\,
\nu _{\mu }(dy) =
 {\frac{2^{\mu }\, \Gamma ({\frac{\mu +1}{2}})}{\pi ^{n/2}|\Gamma (- {\frac{\mu }{2}})|}}
\, \int {\frac{f(y)-f(x)}{|x-y|^{\mu +n}}}\, dy   \label{free1}
\end{equation}
 where $x\in R^n$  and  $\nu _{\mu }(dy)$ stands for a (self-defining) L\'{e}vy measure
  $\sim 1/|y|^{\mu  + 1}$.
  All  above  integrations are  understood in the sense  of the Cauchy principal value
  and $f(x)$ stands for a function in the domain of that (unbounded) operator.

  To set the framework for further discussion, let us mention another (popular, folk) form of
the stable generator in the dimensionless notation:
\be
-|\Delta |^{\mu /2} \equiv {\frac{\partial ^{\mu }}{\partial |x|^{\mu }}}
\ee
and turn over to the  tightly constrained (to remain in a spatial trap)
 L\'{e}vy-stable process, whose transport equation is  typically  \cite{dybiec}-\cite{buldyrev}
  written   in a form mimicking the free motion (quite alike the Brownian case)
\be
\partial _t f(x,t) ={\frac{\partial ^{\mu }}{\partial |x|^{\mu }}} f(x,t)
\ee
even   though  the exterior Dirichlet (absorbing/killing) boundary data are imposed:
$f(x,t) =0$ for    $x\leq a$ and  $x\geq b$, where $a,b \in R$, $t\geq 0$. We have intentionally replaced
 the free motion probability density  function (pdf) $\rho (x,t)$  by more appropriate (spatially nor
 normalized in $L(R)$)  function $f(x,t)$, encoding the killing property. We point out   that random processes with
 absorption/killing   induce their own inventory of  computable quantities,
 like e.g. first exit time, killing time, mean transit time etc. \cite{dybiec}-\cite{lejay}.

 For  the standard (killed) Brownian motion in the interval $D=(a,b) \subset R$, the
 meaning of the locally defined  Laplacian in $\partial _t\rho (x,t) = \Delta \rho (x,t)$,  while
 subject to the very same  exterior  Dirichlet boundary condition $\rho (x,t)=0$ on $R\setminus D$,
$t\geq 0$  is unaffected by the boundary data.  We can evaluate all derivatives of the pdf
 $\rho (x,t)$ point-wise in $D$.

To the contrary, the fractional operator $\partial ^{\mu }/\partial |x|^{\mu }$
cannot  be defined locally.  A  formal  definition (1), if  restricted to  $f(x,t) \neq 0 $  for
 $x\in (a,b)$ only,
is inconsistent. The appropriate functional  form of the constrained  nonlocal operator
 $|\Delta |^{\mu /2} \rightarrow |\Delta |^{\mu /2}_D$, $D=(a,b)\in R$ is definitely lacking in the physics-oriented
literature. For a  mathematical viewpoint on  this issue  see e.g. \cite{KKMS,K}.

   Somewhat disregarded point is that the constrained motion generators determine
   a  corresponding  semigroup dynamics  through a  semigroup kernel  given
   in the familiar Feynman-Kac form.
We shall explore this property in below to pass to nonlocally defined random motions which {\it do }
 preserve a pdf   $\rho $ normalization, while inheriting the constraints (e.g. being trapped in $D$),
 by    following the  line of research developed in Refs. \cite{acta}-\cite{gar}.

\section{What is meant by the Brownian motion in the interval ?}

\subsection{ Schr\"{o}dinger semigroup transcript of the Fokker-Plack dynamics.}

Before embarking on the problem of jump-type processes in a spatially finite trap, let us
invoke a  a classic exercise (\cite{risken}, Section 5.5.3, page
110) illustrating a method of solution of the Fokker-Planck equation
in one variable, by means of so-called eigenfunction expansions. The
major step  there is a transformation of the Fokker-Planck operator
into a Hermitian operator  (which  subsequently needs  to be  elevated to the
self-adjoint or essentially self-adjoint operator status \cite{GK}).
Told otherwise, the Fokker-Planck equation is solved by passing to
an associated Schr\"{o}dinger-type equation. No imaginary unit  appears  here and thence we
 deal not with a unitary evolution, but with the semigroup dynamics.

The essence of the method (we consider the  1D case, in a dimensionless  notation)
lies in passing from the Fokker-Planck equation
\be
\partial _t\rho = \Delta \rho - \nabla \, ( b \cdot \rho ),
 \ee
for the probability density function $\rho (x,t)$,
with the initial condition $\rho_0(x)=\rho (x,0)$ and suitable boundary data,
where  the existence  of  the stationary (equilibrium) pdf   $\rho (x,t) \rightarrow \rho _*(x)$
is presumed to be granted  in the large time asymptotic,
 to the Schr\"{o}dinger-type  equation i.e. the semigroup  $\exp(-Ht)$:
 \be
\partial _t\Psi= - H \Psi  =  \Delta \Psi - {\cal{V}} \Psi \, ,
\ee
for a real-valued    function $\Psi (x,t)$.  We tacitly presume the
potential to be confining so that the  positive definite  ground state  $\psi (x) \doteq \rho ^{1/2}
_*(x)$  exists and corresponds to the $0$ eigenvalue of $H$. This can be always achieved by subtracting
the lowest non-zero eigenvalue  of $H$, if  actually in existence,  from the potential.

The  auxiliary  potential ${\cal{V}}$, up to an additive constant,  takes the form (actually
obeys the  compatibility condition, given $\rho _*(x)$)
\be
{\cal{V}}(x) = \rho _*^{-1/2}  \Delta
\rho _*^{1/2}.
\ee
The transformation between (4) and (5) is executed by means of a substitution (remember that
$\rho (x.t)$, as a probability density function,  integrates to $1$)
\be
\rho (x,t) = \Psi (x,t) \rho _*^{1/2}(x).
\ee
Another  expression  for the  Schr\"{o}dinger  potential  reads $
{\cal{V}} = {\frac{1}2} ({\frac{b^2}{2}}  + \nabla b)$, where $b= \nabla  \ln \rho _*$,
 thus completing the mapping.

Solving (5), with the ground state of $H$ in hands, we readily get a solution  $\rho (x,t)$  of the Fokker-Planck equation
which equilibrates to  $\rho_*(x)$. The technical advantage of the transformation (7) is that we in
 fact can recover a complete spectral solution for $H$, which determines an associated
 semigroup (Feynman-Kac) kernel.  The latter, after accounting for   the ground state $\rho _*^{1/2}$ and the
  so-called Doob's transformation, determines in turn the transition probability density
   of the diffusion  process in question (i.e. that underlying (4)), see  e.g.  Refs.
   \cite{risken,olkiewicz,davies,faris}.
   Another technical advantage is that even quite complicated Fokker-Planck dynamics,
  specifically with no  available   analytic solution, can be addressed by means of  powerful and fairly accurate
   numerical algorithms,  invented for the Schr\"{o}dinger-type (Euclidean, i.e. semigroup) evolution problems,
   \cite{BBC,Auer}.

\subsection{Risken's infinite well.}

In conjunction with the Brownian motion in the interval say
$D= (-1,1)$,  the infinite square well potential,
  with $V(x)=0$ for $x\in (-1,1)\subset R$,  is hereby chosen as a mathematical encoding
   of the Laplacian  with the
  the  Dirichlet  boundary conditions  (so-called  zero exterior condition on $R\setminus D$) imposed on  $L^2([-1,1])$
  functions $\psi (x)$  in its domain:
  $\psi (x) =0$ for $|x| \geq 1$.

    The spectral solution is well known.
In particular we readily have in hands  the lowest
eigenvalue  $\pi ^2 /4$ and the ground state function  $cos(\pi x/2)$ of the  operator   $ -
\Delta $, whose action is  restricted to the well interior.

 The orthonormal
eigenbasis is composed of functions $\psi _n(x)$, $n=1,2,...$ such
that $\psi (x)=0$ for $|x|\leq a$, where  $n$ labels
positive eigenvalues $E_n\sim n^2$. More explicitly: $\psi _n(x)= cos(n\pi x/2)$ for $n$
even and $sin(n\pi x/2)$ for $n$ odd, while the eigenvalues read $E_n = (n\pi /2)^2$.

It is clear that any $\psi \in L^2([-1,1])$, in the domain of the
infinite well Hamiltonian,   may be represented as $\psi  (x) =
\sum_{n=1}^{\infty } c_n \psi _n(x)$.  Its time evolution follows
the Schr\"{o}dinger semigroup pattern $\psi (x)  \rightarrow \Psi
(x,t)=  [\exp(-Ht) \psi ](x) = \sum_{n=1}^{\infty } c_n \exp (-E_n
t)\, \psi _n(x)$

Let us consider $H = -\Delta -  E_1$ instead of $H=-\Delta $  proper (the boundary data being implicit).
Accordingly, the  a priori  positive-definite  ground state $\psi
_1(x) \doteq \rho _*^{1/2}(x)$ corresponds to the zero eigenvalue of
$H- E_1$. Thence, the "renormalized" semigroup evolution reads $\Psi
(x,t) = \exp(+E_1 t) \sum_{n=1}^{\infty } c_n \exp (-E_n t)\, \psi
_n(x) \rightarrow \psi _1 (x) =\rho _*^{1/2}(x)$.   Here, in a
self-explanatory notation  we have  defined the  probability density
function  (pdf)    $|\psi _1(x)|^2= \rho _*(x)$ which is an  equilibrium solution
 of the associated Fokker-Planck equation.

 The semigroup kernel $\exp(-tH)(x,y)$,  associated with such  $H$ whose  lowest eigenvalue is  $0$,
 defines  a time homogeneous random process in the interval.
Its  standard spectral representation  is (the renormalization  by $- E_1$  produces here
 an exponential factor), see also \cite{lejay}:
 \be
 k(t,x,y)= (\exp(-Ht)(x,y) =  \exp (+ \pi^2t /4)  \sum_{n=1}^{\infty } \exp [-(n\pi /2)^2\,  t]\,  \psi _n(x)  \,  \psi _n(y)=
 \ee
 $$
\sum_{n=1}^{\infty } \exp [(1- n^2) \pi ^2\, t/4]  \sin[n\pi (x+1)/2]\, \sin[n\pi (y+1)/2]
 $$
 Here $\Psi (x,t) = \int k(t,x,y) \Psi _0(y) \, dy$.
In probabilistic terms the kernel allows to define  a conditional probability
$P_x(X_t)= k(t,x,y)dy$ that a process started at $x$ will  reach a vicinity $dy$ of $y$ in time $t$.

 In the standard lore of the Brownian motion with killing (sometimes identifed with absorption), one adds that $t$ is prior
 to a killing time $\tau $. An inventory of typical calculable functions/functionals  related to the killed Brownian motion
 (various forms of the transition density $k$, distribution function and density of first exit time $\tau $,
  mean first passage/exit time, etc.) can be found in \cite{redner,borodin,lejay,dybiec}.

\subsection{Fokker-Planck dynamics  in the interval.}

Under the very same infinite well conditions, after taking account of (5)-(7),
 another random process (devoid of any killing notion) is defined by means of the
regular transition probability density (here a multiplicative Doob's transformation is
involved, \cite{olkiewicz}; $x$ and $y$ belong to an open interval  $D$)
\be
p(t,x,y) = k(t,x,y) {\frac{\rho _*^{1/2}(x)}{\rho _*^{1/2}(y)}}
\ee
so that a consistent propagation of the Fokker-Planck  probability density function is secured:
$\rho (x,t)= \int p(t,x,y)\, \rho _0(y)\, dy$ entirely within the interval $D\subset R$.   More details on these and related
issues can be found in \cite{olkiewicz,risken} see also  \cite{faris}.

 The Fokker-Planck equation (4), with a stationary solution $\rho _*(x)$,
  can be rewritten in the form  of the general transport equation
\be
\partial_t\rho = \left[\rho_*^{1/2}\Delta\left(\rho_*^{-1/2}\, \cdot\right)-
\rho_*^{-1/2}\left(\Delta\rho_*^{1/2}\right)\right]\rho.  \label{transport}
\ee
with the  (motion in the interval) boundary data being implicit.\\

{\bf Remark 1:}
This   equation  often  happens to be  explicitly written in terms of $\rho _*(x)= \exp [- \Phi (x)]$ where $\Phi $ plays the
 role of the Boltzmann-Gibbs potential.
In Ref. \cite{SG} we have introduced  a thermal redefinition of  the equilibrium pdf (self-explanatory
notation): $\rho_*(x)= (1/Z)\, \exp [-V(x)/k_BT]$, see also \cite{SG}.\\

{\bf Remark 2:}
The transport equation (\ref{transport}) is amenable to an  immediate generalization to non-Gaussian noises, c.f. \cite{brockmann,SG}.
Anticipating further discussion, we note that a  seemingly "naive"  replacement of $\Delta  $ by a fractional
generator $-|\nabla |^{\mu }$, where $\mu \in (0,2) $ is a stability index, produces the fractional  transport equation  whose dynamics stems form that for the
related fractional semigroup, \cite{SG,brockmann,GS,ZG}. A particular choice of   $\mu =1$ refers to the Cauchy case.
In contrast to the Brownian motion,   the emergent  non-Gaussian (here fractional)  transport equation  cannot be reduced to any known
  Fokker-Planck form, see e.g. \cite{SG,brockmann,klafter} and compare with \cite{fogedby,acta}.
  A proper handling of constraints in the nonlocal random dynamis (e.g. motion in the interval)
  is a subject of our  subsequent  analysis.

\begin{figure}[h]
\begin{center}
\centering
\includegraphics[width=70mm,height=70mm]{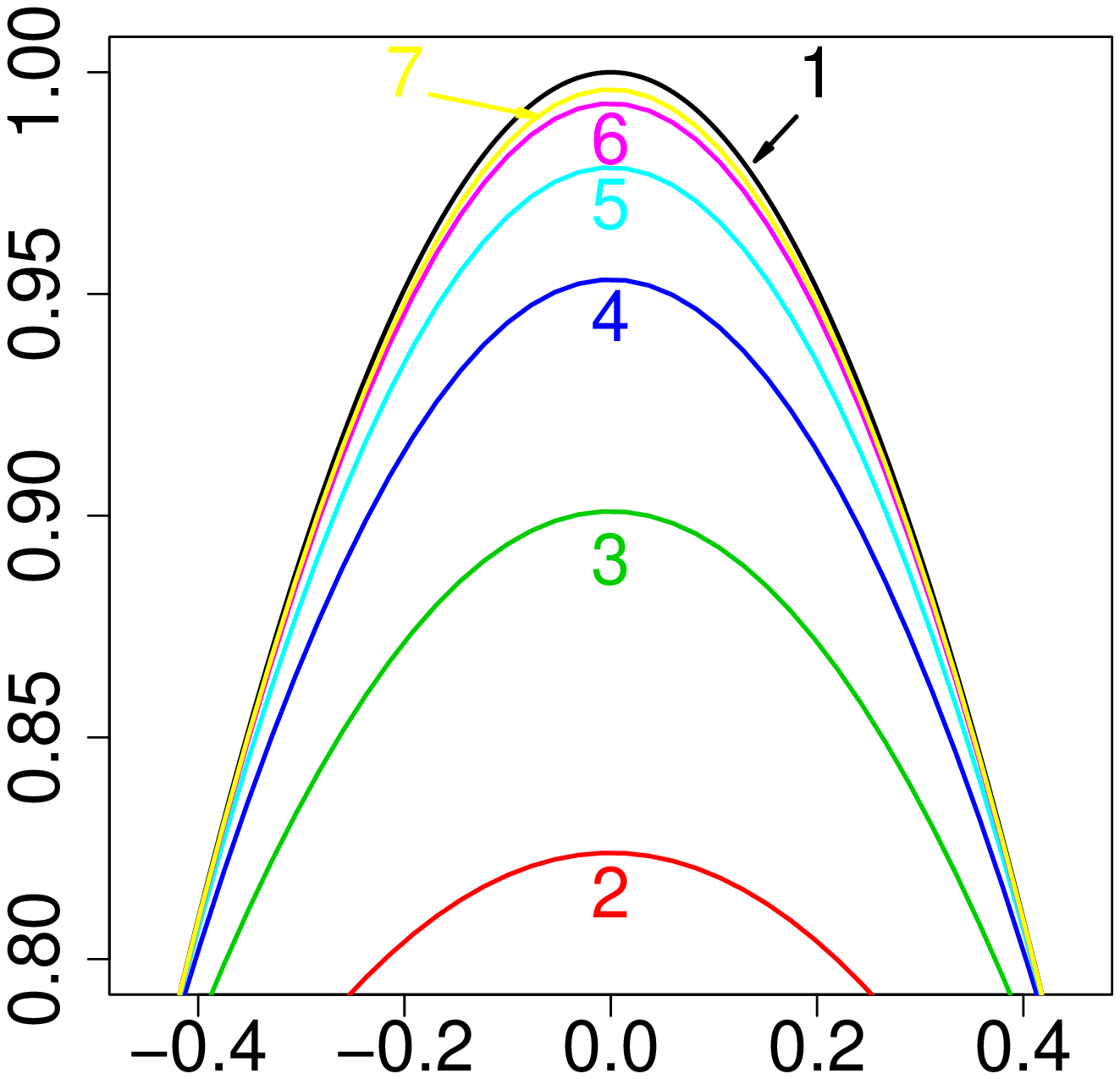}
\includegraphics[width=70mm,height=70mm]{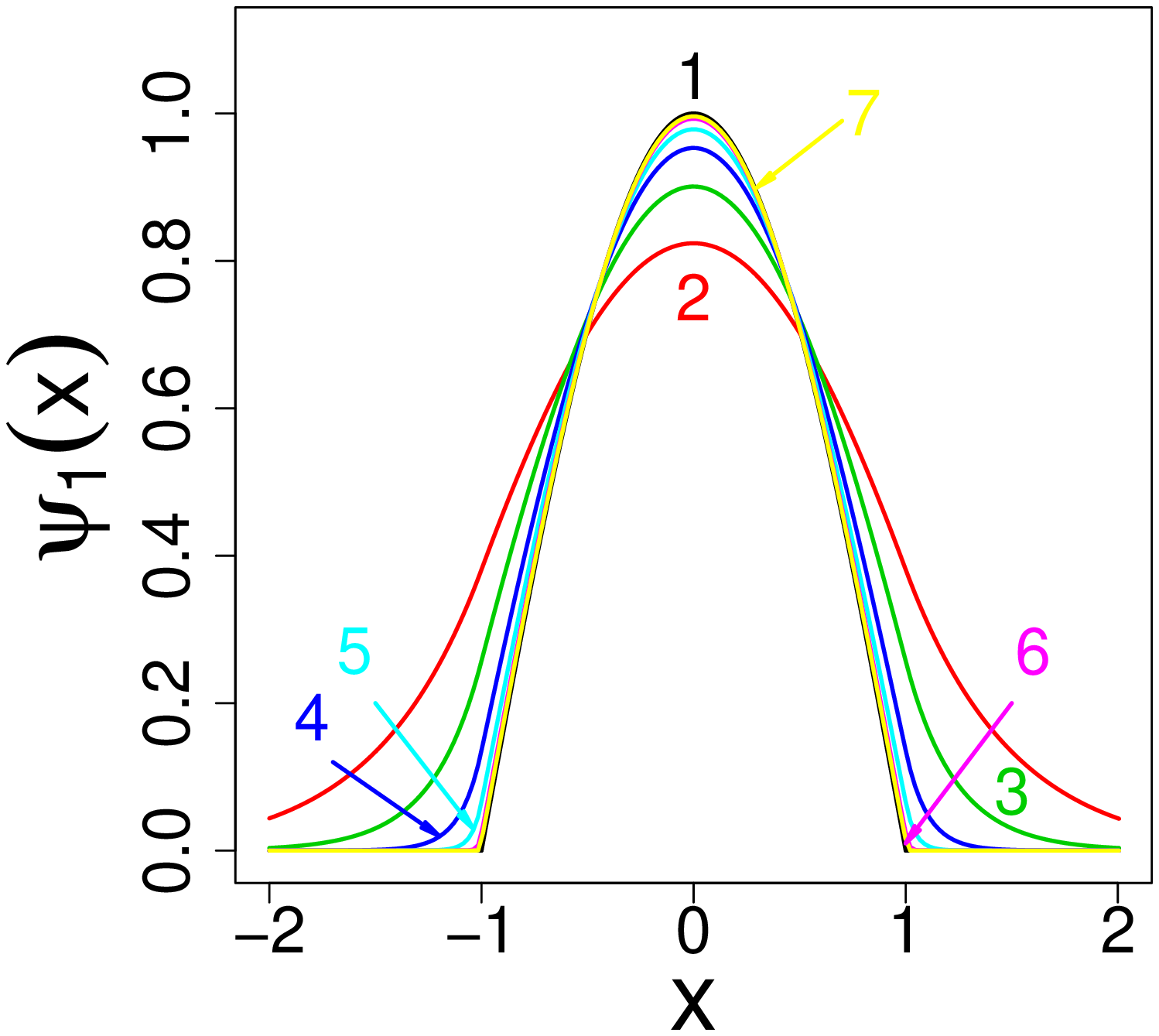}
\caption{Ground states  $\rho _*^{1/2}$  for  a sequence of   deepening  finite wells. Numbers refer to: 1 - $\cos(\pi x/2)$, while  2,3,4,5,6,7  enumerate  well depths
 $V_0=5,20,100,500,5000,50000$ respectively. Left panel shows an enlargement of the vicinity of maxima.}
\end{center}
\end{figure}

\subsection{Relaxing the  boundary data:   Brownian motion in (and in the vicinity of) a finite well}

Following the   previous Risken's recipe, let us consider the Brownian motion in and around the finite well.
We shall be interested in a sequence of deepening  finite wells and the validity of an infinite well
 approximation for wells that are sufficiently deep. By passing to the finite well we relax the "rigid" Dirichlet
 boundary data, since now the pdf tails always persist even far  beyond the well boundaries.

 Let us consider $T=-\Delta $ and  $\cal{V}$ such that  ${\cal{V}}(x)=0$ for $|x|<1$,  while
 ${\cal{V}}(x)=V_0>0$  for $|x|\geqslant 1$.
  The spectral solution  for $H=T + \cal{V}$ is a classic exercise  again. In 1D there exists at least one bound (ground) state and
  a substantial  part of the energy   spectrum  is continuous.

The eigenvalue problem $H\psi(x)=E\psi(x)$  allows to identify at least one (ground state) eigenvalue and the ground state itself.
Namely, the ground state comes from
\be
\psi_0(x)=\left\{
          \begin{array}{ll}
            A\cos(\kappa)e^{k(x+1)}, & \hbox{$x<-1$;} \\
            A\cos(\kappa x), & \hbox{$-1\leqslant x \leqslant 1$;} \\
            A\cos(\kappa)e^{k(1-x)}, & \hbox{$x>1$.}
          \end{array}
        \right.
\ee
where
\be
\kappa=\sqrt{E},\qquad k=\sqrt{V_0-E},\qquad A=\sqrt{\frac{k}{k+1}}.
\ee
and $E$ must be the least real number   obeying
\be
\left\{
  \begin{array}{l}
    k=\kappa\tan(\kappa),  \\
    \kappa^2+k^2=V_0.
  \end{array}
\right.
\ee
The ground state eigenvalues $E=E_1$ for various well depths have been obtained numerically and we reproduce them up to four decimal digits:
\be
\begin{array}{ll}
V_0=5,\qquad &E_1=1.1475,\\
V_0=20,\qquad &E_1=1.6395,\\
V_0=500,\qquad &E_1=2.2605,\\
V_0=1000,\qquad &E_1=2.3184,\\
V_0=5000,\qquad &E_1=2.3989,\\
V_0=50000,\qquad &E_1=2.4296,\\
V_0\sim \infty,\qquad &E_1=  \pi ^2/4 \sim 2.4674.
\end{array}
\ee
We note that   $H-E_1$ has $0$ as the lowest eigenvalue,
the ground state being  identified by setting $E=E_1$ in Eq. (9).

 To avoid the  risk of confusion, we point out that  subsequently   the notation $H$ is  employed for the "renormalized"
Hamiltonian $H-E_1$.
   We denote  $\rho_*^{1/2}=\psi_0$, where $H\rho_*^{1/2}=0$   and,  in accordance with \cite{GS,brockmann}, define  the time evolution  generator in  $\partial_t\rho=L\rho$
 as  $L=-\rho_*^{1/2}H\rho_*^{-1/2}$.   This implies the validity of the  transport equation   (10) which in turn stands  as both (i) the rewriting of
the Fokker-Planck equation and (ii) as the direct consequence  of the semigroup evolution (5), see  e.g. also \cite{SG}.

Functional  shapes  of finite well ground states have been  obtained numerically (details of the algorithm,
inferred from \cite{BBC,Auer}, are  available upon request) and results are depicted in Fig. 1.
 The displayed  finite  well ground states  show  up a conspicuous  (here graphical/visual) convergence trend
 towards  the infinite well ground state   $\cos(\pi x/2)$.
 The ground state  tails  appear to be   relevant  for shallow wells.

\begin{figure}[h]
\begin{center}
\centering
\includegraphics[width=70mm,height=70mm]{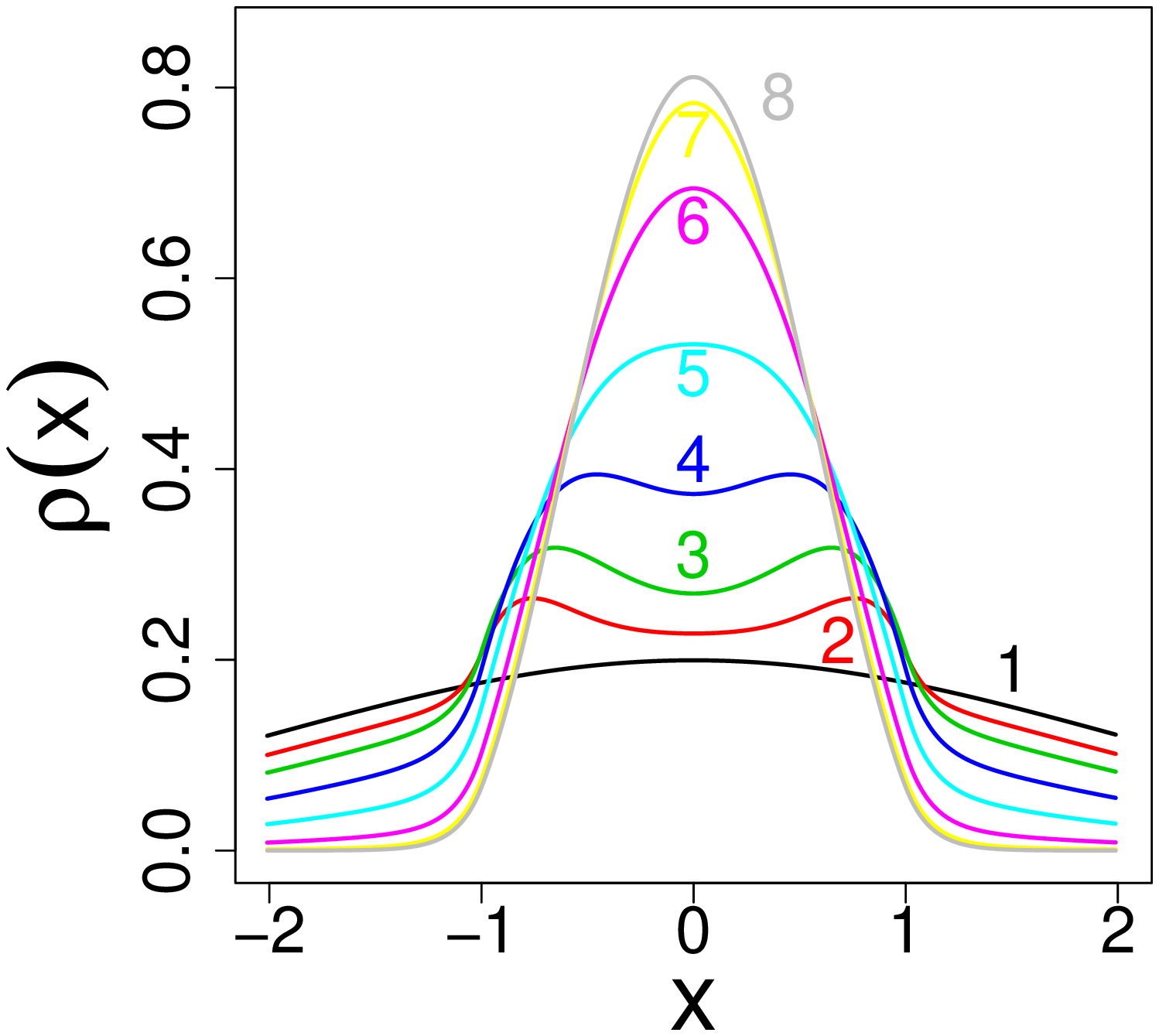}
\includegraphics[width=70mm,height=70mm]{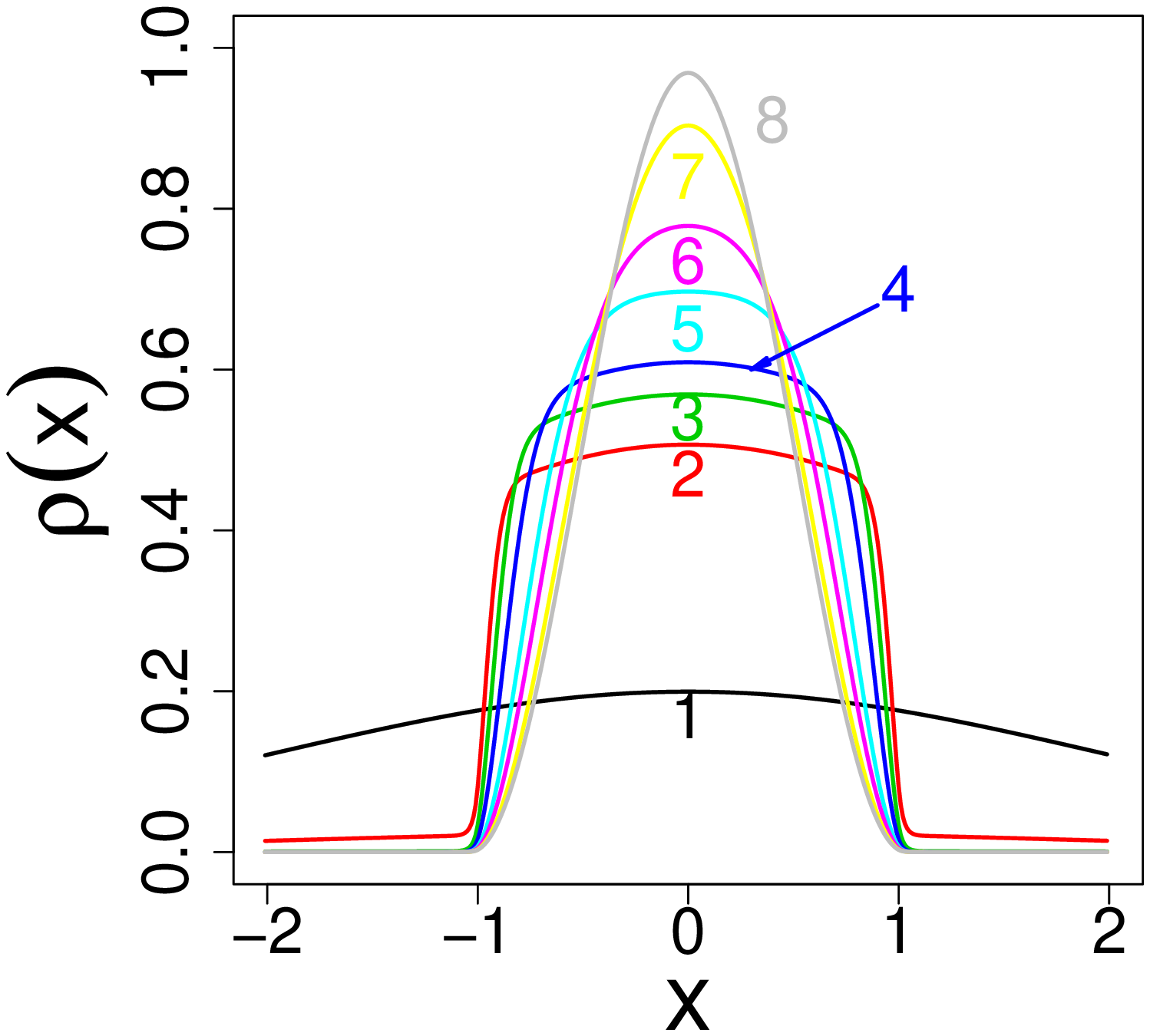}
\caption{Fokker-Planck dynamics of $\rho (x,t)$ in a finite  well   environment.  It is  started from the  gaussian  with cutoffs mentioned in the text.
Numbers refer to: 1 - the gaussian (initial data), $2,3,4,5,6,7$,  depict the $\rho (x,t)$ evolution at selected
 respective  time instants  (number of  algorithm iteration steps). Left panel: for $V_0=20$ we have depicted
time instants $4000, 8000,15 000,25 000,40 000,60 000$, and the vicinity  of an asymptotic ($8$) for  $120 000$.
Right panel: for $V_0=1000$ the evolution proceeds   somewhat  faster and we have respectively
 $300, 600, 1200, 3000, 5000, 10000$ while  $8$  refers to  $100 000$.  The time increment equals  $\Delta t = 10^{-5}$}
\end{center}
\end{figure}

Let us recall that, in accordance with (1)-(4),  given the ground
state of (2) and the semigroup-driven evolution of $\Psi (x,t)$, we have readily defined
   the pdf of  the   Brownian motion without killing, albeit asymptotically (almost)    trapped   in the finite well.
   Indeed, we have $  \rho (x,t) = \Psi (x,t) \rho _*^{1/2}(x)$ and  the time evolution  of  $\rho (x,t)$
    is fully compatible with the   Fokker-Planck equation (1).  To visualize the finite well dynamics in its Fokker-Planck transcript,   let us consider  the
     standard gaussian as an initial pdf: $ \rho_0(x)=\frac{1}{\sigma\sqrt{2\pi}}\exp(-(x-\mu)^2/2\sigma^2)$,
      where  we set  $\mu=0, \sigma=2$.\\

{\bf Remark 3:}  The transport equations (4) and (10) can be rewritten as  $\partial _t \rho = L\rho $
where the operator  $L$  reads: $L= -\rho _*^{1/2} \, H\, \rho _*^{-_1/2}$.  The dynamics of $\rho (z,t)$ can
be simulated by employing the standard finite difference scheme. Namely, for a sufficiently small time
increment  $\Delta t$, we can set  $\rho (x,t+\Delta t) \approx \rho (x,t) + \Delta t \,
 (L\rho )(x,t)$. In the Brownian case  $\Delta t = 10^{-5}$ is a reliable choice.
 After each simulation step the outcome needs to be  normalized to yield a consistent approximation
  $\rho (x, k\Delta t)$, $k=0,1,2,...$ of the probability density function at the time instant $k\Delta t$.\\

{\bf Remark 4:}   To facilitate  numerical computations (optimize the time necessary to get close to the  equilibrium pdf),
  in case of wells with  $V_0=500$ and $V_0=1000$  the
 support of the gaussian is restricted to $[-10,10]$ which is  followed by the $L^2(R)$ normalization of the outcome.  In case  of shallow
wells, $[-a,a]$ with $a=50$ has been employed for $V_0= 20$, while $a=150$ for  $V_0=5$.
Clearly,  time  (in terms of the computer algorithm, it is the  number of steps) necessary to
reach the vicinity of an equilibrium is considerably longer for shallow wells.

\section{Cauchy process in (and in the vicinity of) the   finite well.}
\subsection{Finite Cauchy wells.}

\begin{figure}[h]
\begin{center}
\centering
\includegraphics[width=75mm,height=75mm]{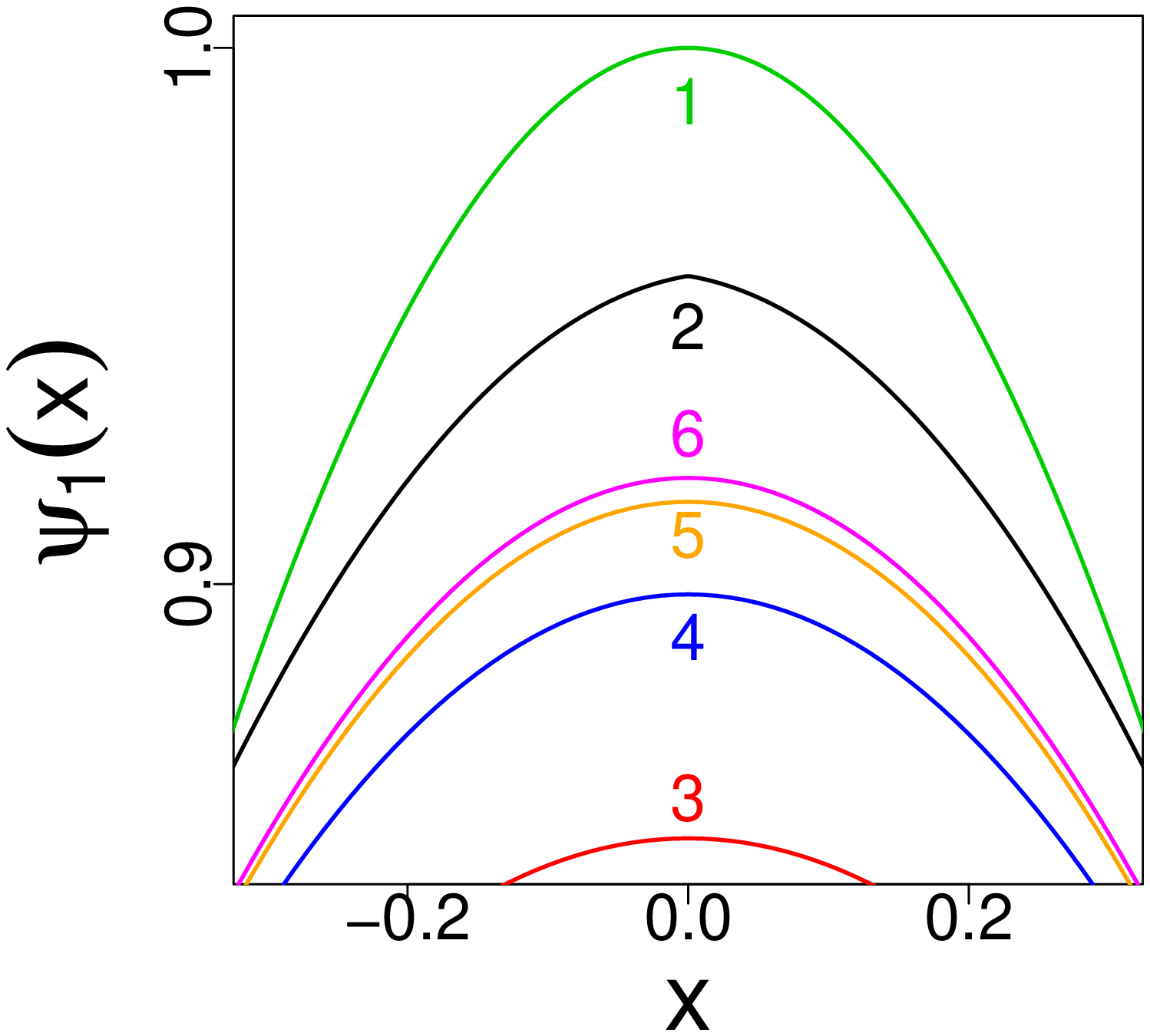}
\includegraphics[width=75mm,height=75mm]{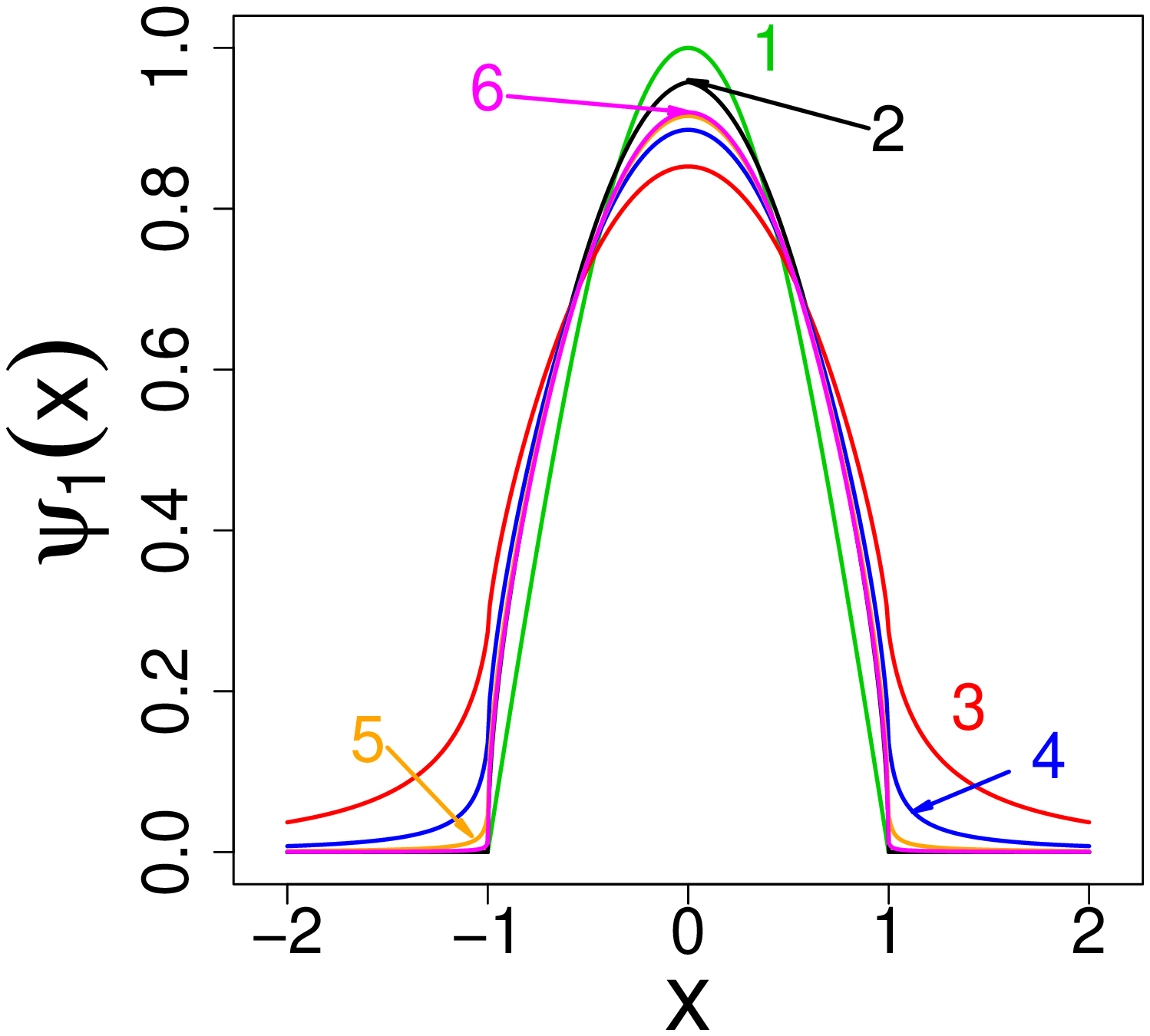}
\caption{Ground state solution of  the  finite Cauchy  well. Numbers refer to: 1 - $\cos(\pi x/2)$, 2 - an approximate solution, Eq. (13) in \cite{K}, 3,4,5,6
  refer to the  well depths, respectively  $5, 20, 100, 500$. Convergence symptoms (towards an infinite well solution) are visually identifiable.
Left  panel reproduces an  enlarged  resolution around the maximum of the  ground state.}
\end{center}
\end{figure}

 Our major point of interest  is an exemplary nonlocal  Cauchy  jump-type process, that is
 constrained  to "live" exclusively within the interval $(-1,1) \subset R$,  due to killing
(absorption) at the boundaries, according to the traditional probabilistic lore, \cite{K}-\cite{ZG}.
In the mathematical literature the process is considered in the finite interval, as a problem for itself.
We wish to maintain at least a residual link with physical intuitions about trapping mechanism. In
particular, it may be illuminating to have in hands a model which shows how  long jumps can be   tamed,
with an ultimate reduction of their impact once  we approach the  motion  restricted to the  interval only.

To this end let us  introduce   somewhat  milder boundary conditions (permitting
 arbitrarily long jumps to occur; note that in numerical procedures their length needs to bounded by a certain $a>0$),
by referring to  a sequence of deepening but  finite Cauchy  wells.
Spectral solutions for  sufficiently deep   wells
may be satisfactorily   approximated by that for  the infinite one, see e.g. \cite{K}.
 For finite wells, there are  no a priori limitations upon  the size of jumps in the
pertinent  jump-type process and the  $R$-nonlocality of the problem persists,  while for the
infinite well its impact is limited to the interior of $D\subset R$.

We have in hands    \cite{ZG} numerical tools allowing to   deduce the corresponding   ground states  (actually approximants of the "true" ones)
and  the semigroup  dynamics of $\Psi (x,t)$  in the finite well regime.
 A  family of  related  pdfs  $\rho (x,t); \, t\geq 0$
  readily follows by employing the  transformation (7).     The inferred    probability density functions (pdfs)
 are  driven towards equilibrium by a   suitable master equation. We point out that, in contrast to the Brownian motion of Section II,
this equation cannot be reduced to any known Langevin-based form of the fractional Fokker-Planck equation, \cite{brockmann,SG,klafter,fogedby}.

 In the  finite well (semigroup) regime, a family of jump-type processes running in sufficiently deep wells, admits only   the residual
tails of the equilibrium pdf  to persist beyond the trap (e.g. $D=(-1,1)$)
interior. A degree of an approximation accuracy,   with which the infinite well data are reproduced,   is quantified in
terms of deviations   of  each  equilibrium pdf    from that emerging in
the reference $(-1,1)\subset R$ trapping model. This  model  we have investigated before, \cite{ZG}
 see  also \cite{K}.  It is the complete   "blockade"  of long jumps, that  ultimately
  calls for  giving a  meaning  to the
Dirichlet-constrained   nonlocal   generator $|\nabla |_D$, where  $D=(-1,1)$.

We consider  the  Cauchy-Schr\"{o}dinger semigroup dynamics  $\exp(-Ht)$     where  $H= T+V- E_1$ and  $-T$
stands for the Cauchy generator, e.g.   $T= |\nabla | = (-\Delta )^{1/2}$, while  $V$ denotes the finite well potential defined in Section
 II.D and $E_1$ is the bottom
(ground state) eigenvalue of $H$. Here
\be
T\,\psi(x) = (-\Delta)^{1/2}\,\psi(x)=\frac{1}{\pi}\int\frac{\psi(x)-\psi(x+z)}{z^2}dz,
\ee
and the integral is interpreted in terms of the Cauchy principal value.

The semigroup evolution gives rise to the transport equation for $\rho (x,t) = \Psi (x,t) \rho _*^{1/2}(x)$,
 which is a straightforward generalization of Eq. (10) mentioned in Remark 2, see for more details
 \cite{acta,SG,GS}:
\be
\partial_t\rho = -\left[\rho_*^{1/2}T\,\left(\rho_*^{-1/2}\cdot\right)-\rho_*^{-1/2}\left(T\rho_*^{1/2}\right)\right]\rho.
\ee
where $\rho_*^{1/2}$  is the $L^2(R)$ normalized ground state of  $H=T+V- E_1$, associated
 with the eigenvalue $0$.

\begin{figure}[h]
\begin{center}
\centering
\includegraphics[width=70mm,height=70mm]{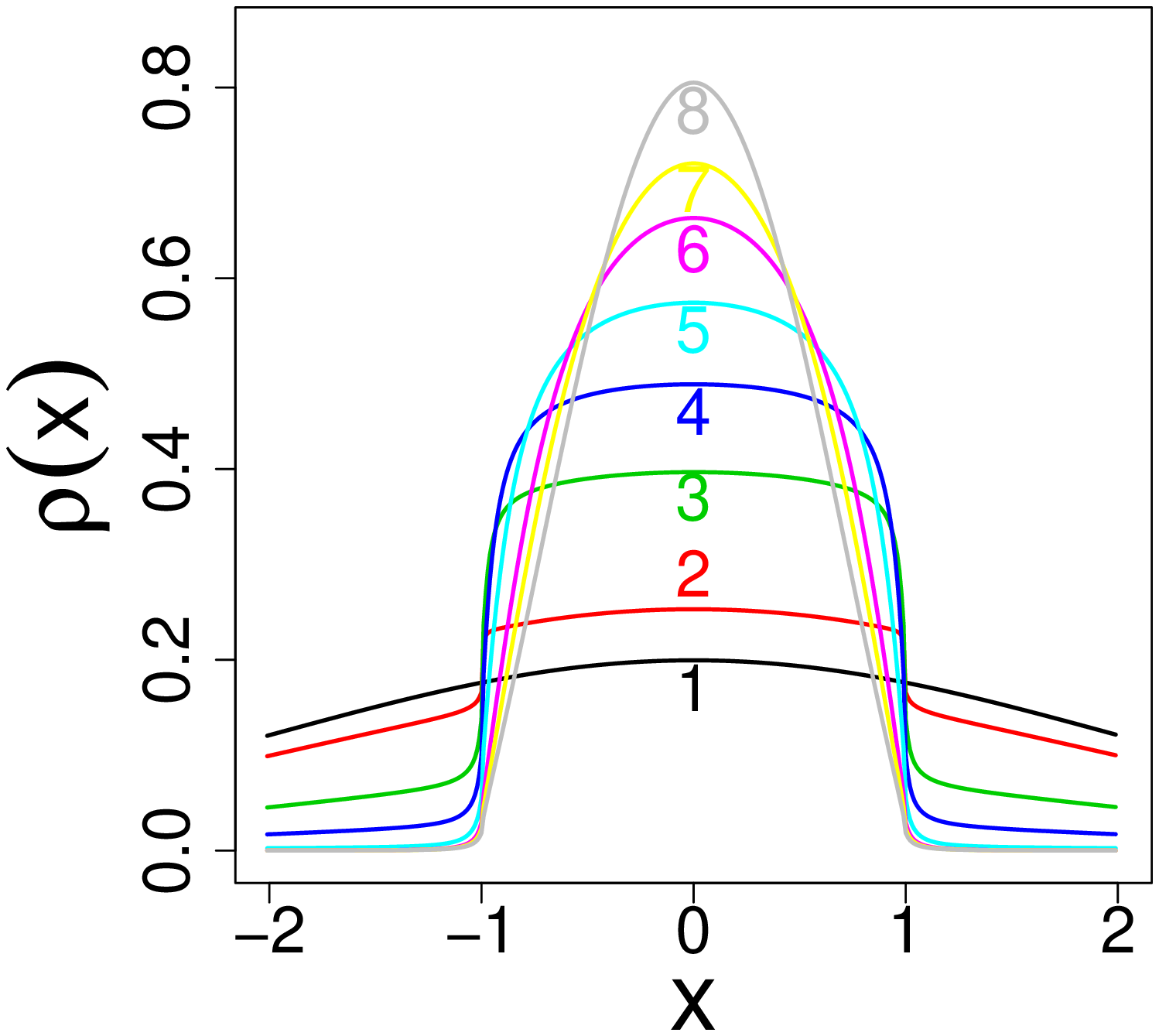}
\includegraphics[width=70mm,height=70mm]{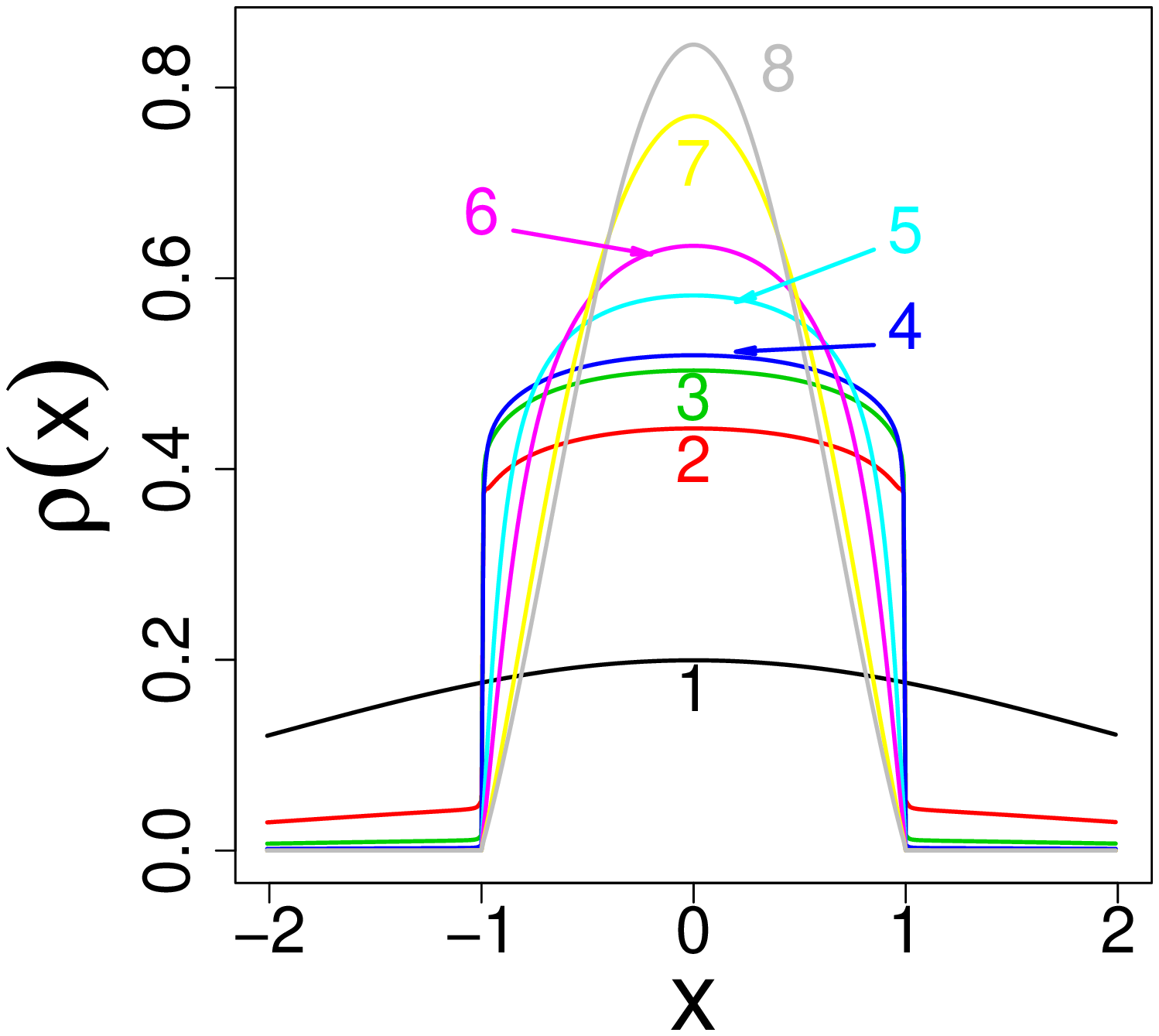}
\caption{Cauchy evolution of $\rho (x,t)$  in the  finite well environment.  Left panel $V_0=20$:
numbers refer to: 1 - initial gaussian pdf, $2,3,4,5,6,7$,   algorithmic time instants after  $10, 50, 100, 200, 400, 600$    steps,
 8 - a close vicinity of an asymptotic pdf is approached after  $2500$ steps. Right panel  $V_0=500$:
 1 - initial gaussian pdf, $2,3,4,5,6,7$, refer yo  $2, 4, 6, 100, 200, 600$   algorithm steps respectively, 8 - a vicinity of an
asymptotic pdf after  $2000$ steps. Time increment $\Delta t = 10^{-3}$ is $100$ times  larger than that adopted for Brownian simulations.}
\end{center}
\end{figure}

\begin{figure}[h]
\begin{center}
\centering
\includegraphics[width=70mm,height=70mm]{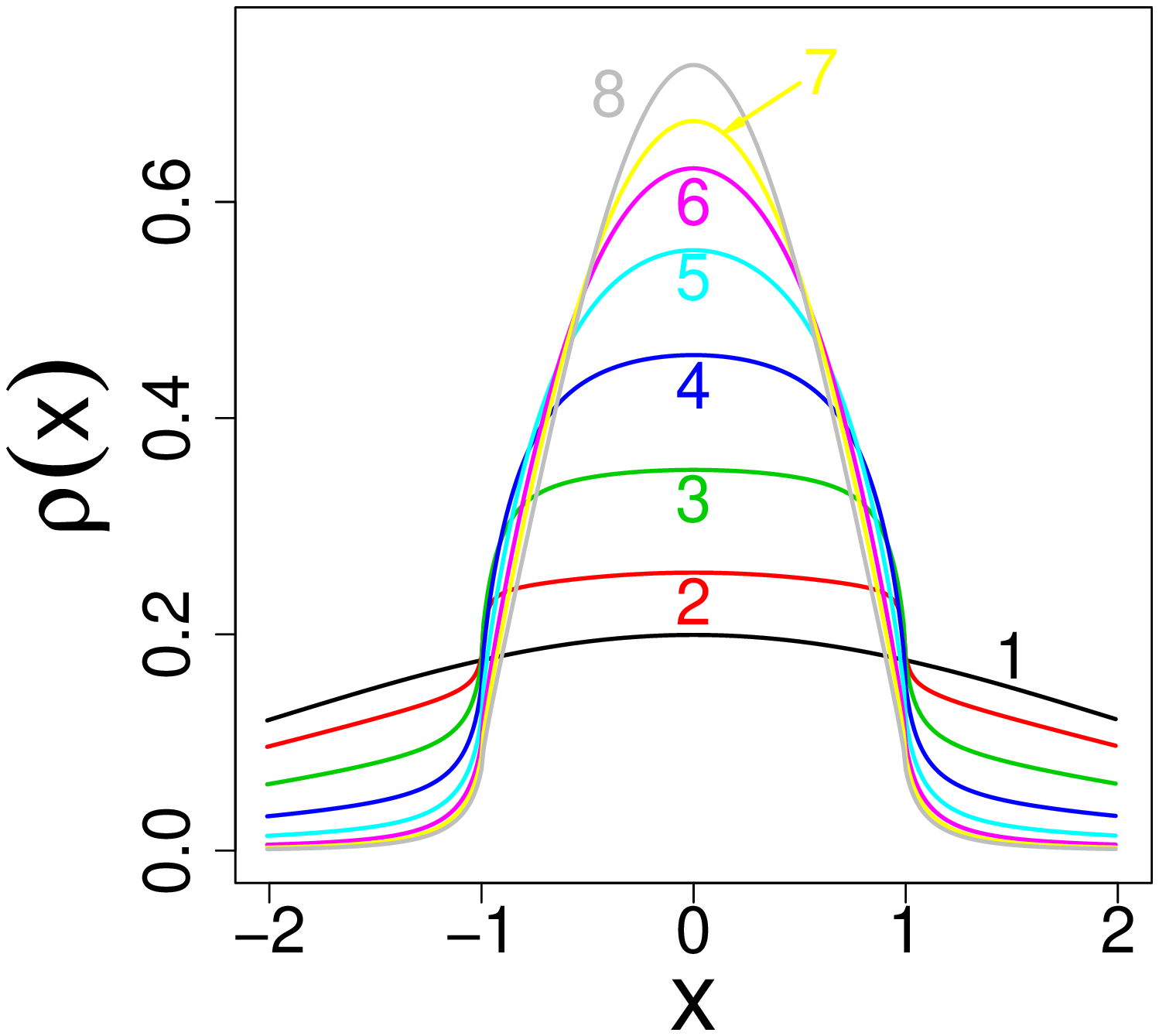}
\includegraphics[width=70mm,height=70mm]{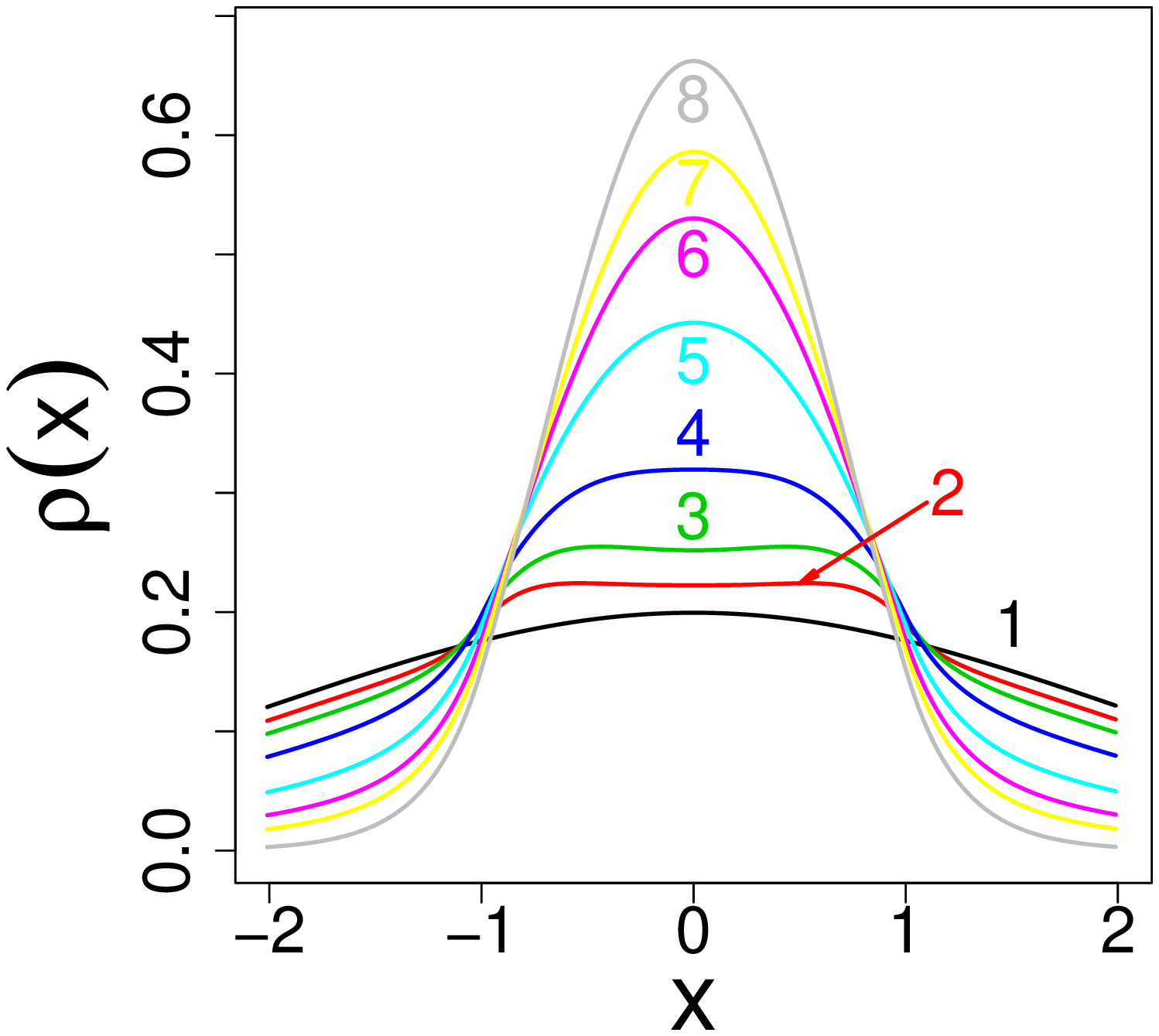}
\caption{Cauchy evolution versus Brownian evolution  of $\rho (x,t)$  in the  finite well environment,  $V_0=5$:
Left panel: Cauchy driver,  1 - initial gaussian pdf,  $2,3,4,5,6,7$,  refer to algorithmic time instants set by  $50, 150,
300, 500, 750, 1000$ steps, while  $8$ to an asymptotic retrieved after  $3000$ steps.  Here $\Delta t = 10^{-3}$.  Right panel: Brownian driver,
1 - initial gaussian pdf,  $2,3,4,5,6,7$ refer  to algorithmic   time instants set by   $5000,10 000,20 000, 40 000, 60 000$ steps, with an asymptotic
$8$ approached after  $150 000$ steps.  Here $\Delta t = 10^{-5}$.}
\end{center}
\end{figure}

The finite Cauchy well ground  (and higher energy)  states we have numerically recovered   for various wells depths in Ref. \cite{ZG}. A
suitable  algorithm  (based on the Strang splitting method) has been implemented    there
  for the semigroup evolution of $\Psi (x,t)$, including an issue of its large time  asymptotic $\rho _*^{1/2}$.
As a consequence,  we  can readily  deduce   the evolution of the  inferred  pdf
 $\rho (x,t)=\Psi (x,t) \rho _*^{1/2}(x)$ whose asymptotic $\rho _*(x)$ actually is.

We recall  \cite{brockmann,SG,gar} that the transport equation (16) cannot be reduced to any
 traditional  form of  the Langevin-based  fractional Fokker-Planck equation, like those discussed
 in Refs. \cite{fogedby,klafter}.

Analytic outcomes are generically  beyond the reach and the computer assistance is unavoidable in the present context.
Various technical details of developed  numerical routines are skipped in the present paper, see e.g. \cite{ZG}.
It is useful to mention that integrations involved in the definition of the Cauchy generator need to be chosen finite,  to optimize
 computations.
 In view of the preselected  trap size $D=(-1,1)$, a reliable cutoff for the integration domain is  $x\in[-a,a]$, with  $a=50$.
We have analyzed before \cite{ZG}  a sensitivity of computed eigenvalues (high) and   eigenfunction shapes (low) on the cutoff parameter $a$.

In the Cauchy case we set  a time increment $\Delta t = 10^{-3}$, which is $100$ times longer  than that adopted for the Brownian case. While attempting
any comparison of  Brownian and Cauchy outcomes, one needs to remember that e.g. $100$ Cauchy algorithmic steps corresponds to time
$10^{-1}$ and this time  instant in turn does correspond to $10 000$ algorithmic steps in the Brownian case.

Figs. (4) and (5)  provide a  visualization, in terms of $\rho (x,t)$   that has been started from a gaussian,   of how the entrapping of the Cauchy
process takes place in the $(-1,1)$ finite well environment. We have depicted both shallow ($V_0= 5, \, 20$) and deep ($V_0=500$) wells.
The term "asymptotic" refers to
time regimes such that the resultant  "asymptotic  curve" cannot be distinguished from an independently obtained stationary
 $\rho _*(x)$. At least within the  adopted  graphical resolution.

In Fig. (6) we compare Cauchy and Brownian trapping scenarios in a shallow well  $V_0=5$ environment.

\subsection{Infinite Cauchy well:  ground state function problem.}

\begin{figure}[h]
\begin{center}
\centering
\includegraphics[width=75mm,height=75mm]{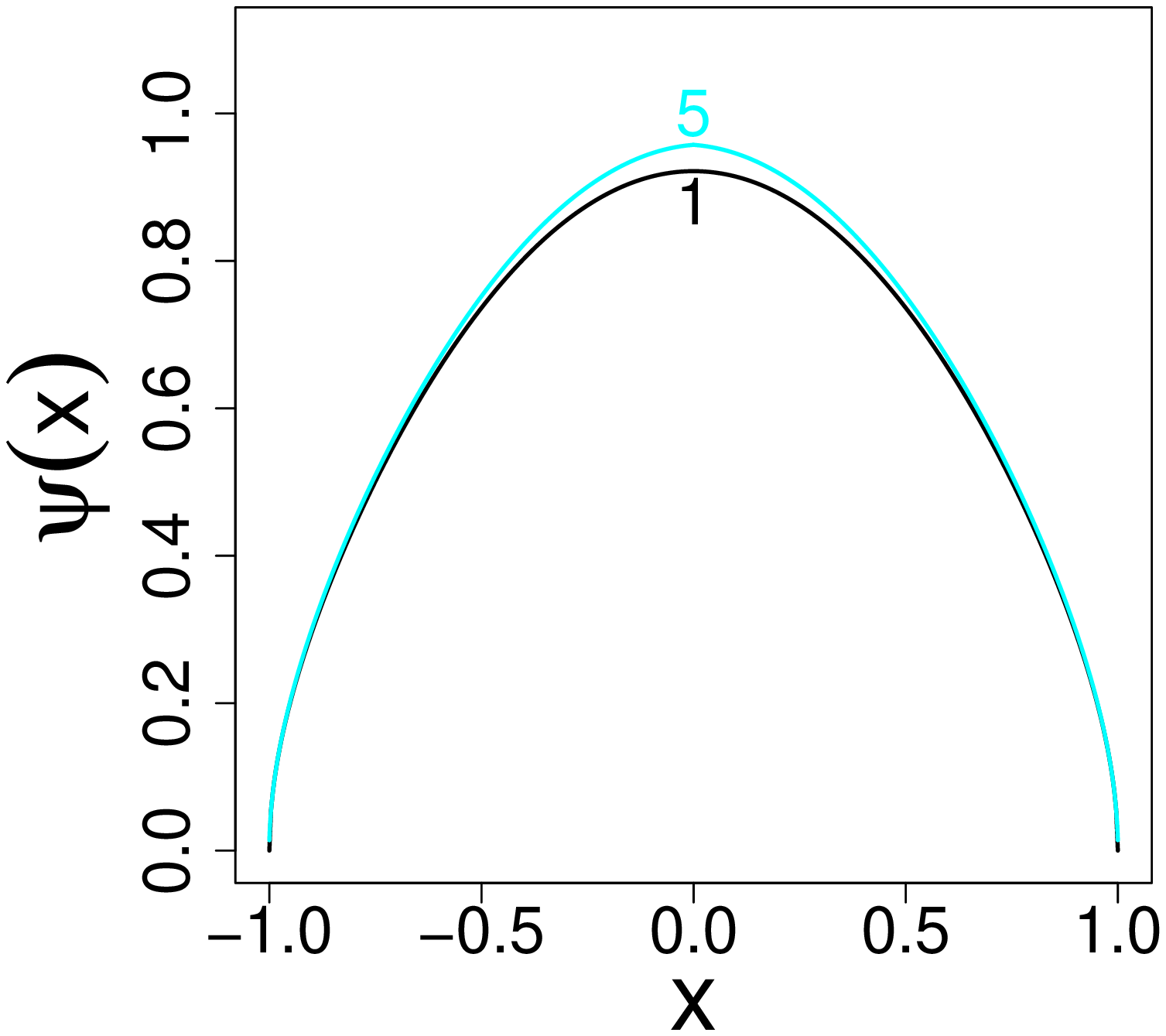}
\includegraphics[width=75mm,height=75mm]{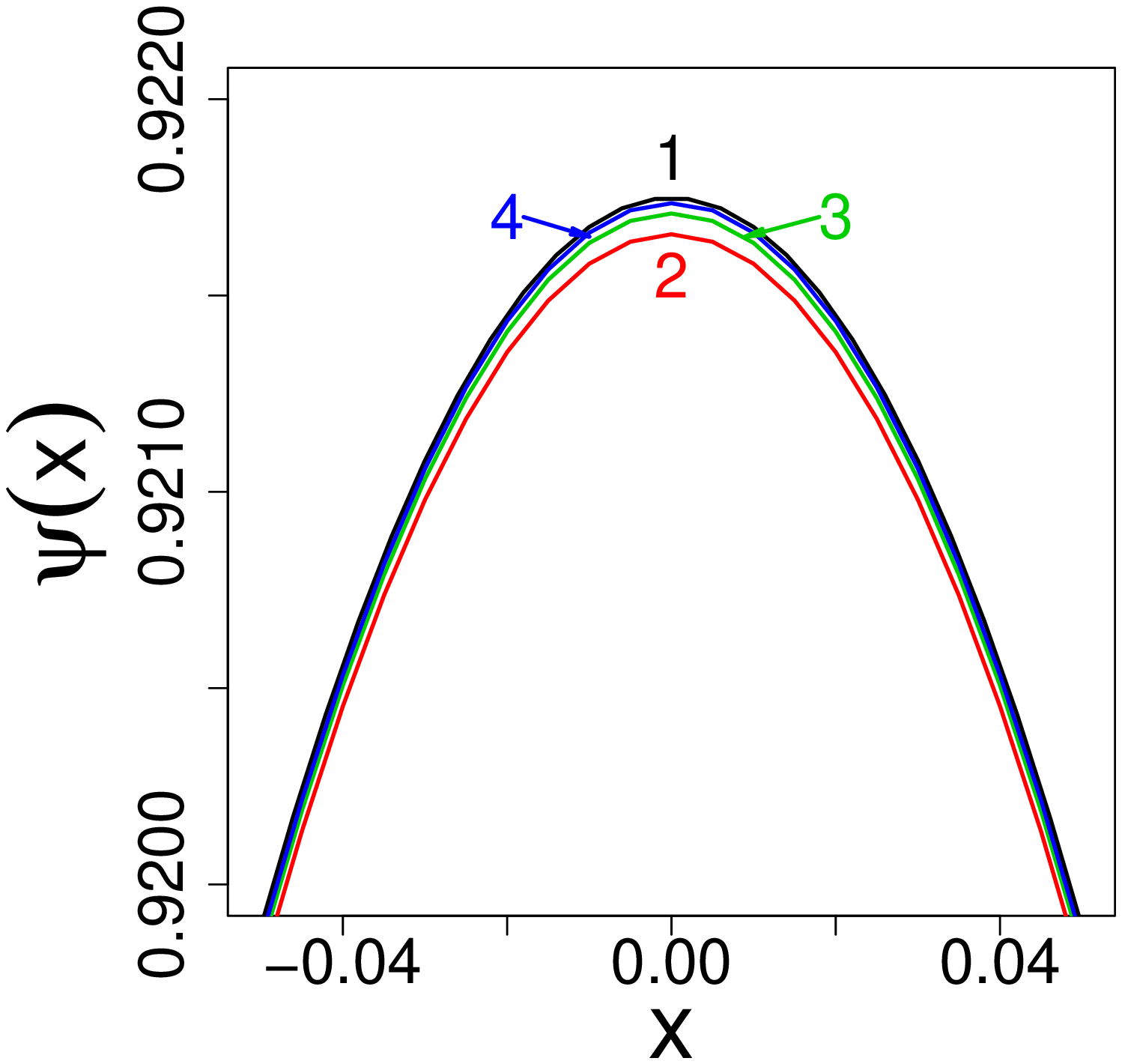}
\caption{Ground states for deep  Cauchy wells. Numbers denote: $1$- our analytic proposal  $\psi(x)$ for the infinite well, 2,3,4 -  well depths
 $V_0=5000,10000,20000$  respectively, 5 - an  aproximation of the infinite well ground state proposed in Ref. \cite{K}.
 Notice that in the   right panel, where an enlargement around the maxima is depicted,  the approximating curve 5 could not be fit to the current
 panel area, in view of  adopted fine resolution  scales.}
\end{center}
\end{figure}

   \begin{figure}[h]
   \begin{center}
   \centering
   \includegraphics[width=75mm,height=75mm]{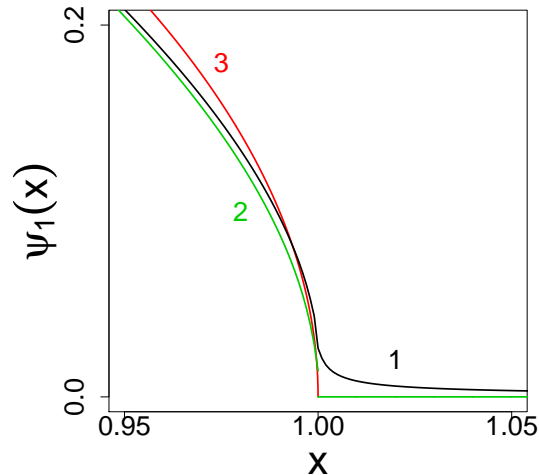}
   \caption{Finite versus  infinite  Cauchy  well ground state  in the vicinity of  the boundary $+1$ of $[-1,1]$: 1 (black) - the  algorithm   outcome for the finite  $V_0=500$  well,
    2 (green)  - an  approximate infinite well  expression from Ref. \cite{K}, 3 (red) -  an approximate form  of
   $\psi _1(x)\sim (1-|x|)^{1/2}$, in the vicinity of the  infinite well barriers,  as proposed in  Ref. \cite{ZRK}.}
   \end{center}
   \end{figure}

In case of  the Cauchy well,  our numerical  algorithm  (i)-(v)  allows to  deduce  approximate eigenvalues and eigenfunctions of the  pertinent  spectral problem.
 Since an  infinite well limit, to which we continually refer, can be formulated  {\it solely} in terms of $L^2([-a,a])$ (see however \cite{GK} for another viewpoint in the local context),
 we  may  quite intentionally  consider   an   orthonormal basis  in
 $L^2[-1,1]$, subject to a trivial  extension  to $L^2(R)$:
\be
\Phi_{n=2m+1}^{(0)}(x)=\left\{
                    \begin{array}{ll}
                      A\cos\left(\frac{n\pi x}{2}\right), & \hbox{$|x|<1$,} \\
                      0, & \hbox{$|x|\geqslant 1$}
                    \end{array}
                  \right.\qquad
\Phi_{n=2m}^{(0)}(x)=\left\{
                    \begin{array}{ll}
                      A\sin\left(\frac{n\pi x}{2}\right), & \hbox{$|x|<1$,} \\
                      0, & \hbox{$|x|\geqslant 1$}
                    \end{array}
                  \right.\qquad
m=0,1,\ldots
\ee
The normalization constant $A$  equals  $\pm 1$.  Generically, a particular sign choice  seems to have  no physical meaning.
However, in view of our semigroup discussion we adopt $A=1$ which secures that an asymptotic $\Psi (x,t)\rightarrow \rho _*(x)$ has
an unambiguous meaning.
The  above   trigonometric  functions actually stand for a complete set of eigenfunctions of the Brownian infinite well problem. \\

In the fractional (like e.g. Cauchy) context many authors have claimed that trigonometric functions should be close to the "true" eigenfunctions
of  fractional semigroups in the interval. See e.g.  \cite{SG,ZG,K}  for  a brief summary of statements and pitfalls related to this issue.
One can possibly  accept a statement that for sufficiently large $n$ the eigenvalues  are close to $E_n = n\pi /2$ and eigenfunctions  are close to
the above trigonometric basis system  in $L^2([-1,1])$.

As yet no explicit analytic formula for any fractional semigroup  ground state in the infinite well is available.
Approximate analytic expressions of a  relatively good finesse are known.This statement extends to the eigenvalues as well.
Coming back to the Cauchy infinite well problem  with boundaries at  the ends of  $[-1,1]$, we note that for low lying eigenvalues
the formula  $E_n = n\pi /2$ is plainly wrong. Actually we have  (it is still an analytic approximation,  provided $n$ is not too
small - in fact $n>10$ does the job), \cite{K}:
\be
 E_n = {\frac{n\pi }2}  - {\frac{\pi}8} + O\left({\frac{1}{n}}\right).
 \ee
In \cite{ZG} we have deduced numerically  the Cauchy infinite well eigenvalues,
 together with  shapes of the corresponding  eigenfunctions. The level of accuracy for low lying
 eigenstates is surprisingly good.

For completeness of arguments,  let us  give an explicit expression for approximate eigenfuctions
 associated with the infinite Cauchy   well.
Namely, we have (with minor adjustments of the original notation  of Ref. \cite{K}):
\be
\psi_n(x)= q(-x)F_n(1+x)-(-1)^nq(x)F_n(1-x),\qquad x\in R,
\ee
where  $E_n=\frac{n\pi}{2}-\frac{\pi}{8}$ and  $q(x)$   is an auxiliary function
\be
q(x)=\left\{
       \begin{array}{ll}
         0 & \hbox{for $x\in(-\infty,-\frac{1}{3})$,} \\
         \frac{9}{2}\left(x+\frac{1}{3}\right)^2 & \hbox{for $x\in(-\frac{1}{3},0)$,} \\
         1-\frac{9}{2}\left(x-\frac{1}{3}\right)^2 & \hbox{for $x\in(0,\frac{1}{3})$,} \\
         1 & \hbox{for $x\in(\frac{1}{3},\infty)$.}
       \end{array}
     \right.
\ee
The function $F_n(x)$  is defined as   follows: $
F_n(x)=\sin\left( E_n\,  x+\frac{\pi}{8}\right)-G(E_n \,  x)$,
where  $G(x)$  is the Laplace transform $
G(x)=\int\limits_0^\infty e^{-x s}\gamma(s) ds $ of a positive definite  function $\gamma(s)$:
\be
\gamma(s)=\frac{1}{\pi\sqrt{2}}\frac{s}{1+s^2}\exp\left(-\frac{1}{\pi}\int\limits_0^\infty\frac{1}{1+r^2}
\log(1+r s)dr\right).
\ee
Evidently, things are here much more  complicated than  an oversimplified (deceiving but faulty)  guess (16) would suggest.

Since it is the ground state that matters in our discussion of the inferred pdf $\rho (x,t)$ dynamics, let us  introduce
another analytic approximation of the "true"  ground state in the Cauchy case.
Namely, while skipping a number of detailed hints that motivate our choice,
 we   propose  the following function as the pertinent  approximation
\be
\psi(x)=C\sqrt{(1-x^2)\cos(\alpha x)},
\ee
where
\be
\alpha=\frac{1443}{4096}\pi=(\frac{\pi}{2}-\frac{\pi}{8})-\frac{\pi}{64}-\frac{\pi}{256}-\frac{\pi}{512}-\frac{\pi}{1024}-\frac{\pi}{4096},
\ee
and   $C=0.921749$  is a normalization constant.  We note that the boundary behavior of our $\psi $ conforms with that predicted
 by  means of  scaling arguments in \cite{ZRK}, e.g. drops down to $0$ as  $(1-|x|)^{1/2}$.  Clearly, $\psi $
 becomes  close to the cosine once away from the boundaries of $[-1.1]$.

The function is concave and conforms with earlier mathematical results on the  the ground state shape for
 stable  generators in the interval, \cite{B,B1}.
 It is somewhat funny to note that $\psi ^2(x)= \rho _*(x)$, up to an overall normalization, comprises  a product of the cosine and
Wigner semicircle distributions.

The approximation accuracy with which our $\psi (x)$ mimics the numerically obtained  very
deep  (and ultimately the infnite)  Cauchy well ground states seems
to be better than that offered by the analytic proposal of \cite{K}, see e.g. Figs. (6) and (7).


\begin{thebibliography}{50}
\bibitem{dybiec} B. Dybiec, E. Gudowska-Nowak and P. H\"{a}nggi,
\textit{L\'{e}vy-Brownian motion on finite intervals: Mean first
passage time analysis}, Phys. Rev. E{\bf 73}, 046104, (2006)
\bibitem{ZRK}  A. Zoia, A. Rosso and M. Kardar, \textit{Fractional Laplacian in a bounded domain},
 Phys. Rev. E \textbf{76}, 021116,(2007).
\bibitem{drysdale} P. M. Drysdale and  P. A. Robinson,
\textit{L\'{e}vy random walks in finite systems}, Phys. Rev. E{\bf 58}, 5382, (1996)
\bibitem{buldyrev} S. V. Buldyrev et al, \textit{Properties of L\'{e}vy flights on
an interval with absorbing boundaries},  Physica {\bf A 302}, 148, (2001)
\bibitem{faris}  W. G. Faris, \textit{Diffusive motion and where it leads}, in:   W. G. Faris (Ed.),
\textit{Diffusion, Quantum Theory and Radically Elementary Mathematics}, (Princeton
University Press, Princeton, 2006)
\bibitem{davies} E. B. Davies, \textit{Heat kernels and spectral
theory}, (Cambridge University Press, Cambridge, 1990)
\bibitem{redner} S. Redner, \textit{A guide to  first-passage processes}, (Cambridge University Press, Cambridge, 2001)
\bibitem{borodin} A. N. Borodin and P. Salminen, \textit{Handbook of Brownian Motion - Facts and Formulae},
(Birkh\"{a}user, Basel, 2002)
\bibitem{lejay} A. Lejay, \textit{A library of simulating Brownian motion's exit times and positions from simple
 domains}, Rapport technique No 7523, (INRIA, Nancy, 2011)
\bibitem{KKMS} T. Kulczycki, M. Kwa\'{s}nicki, J. Ma{\l}ecki and  A. St\'{o}s, \textit{Spectral properties of
the Cauchy process on half-line and interval}, Proc. London Math. Soc. 101, 589, (2010).
\bibitem{K} M. Kwa\'{s}nicki, \textit{Eigenvalues of the fractional Laplace operator in the interval},
J. Funct. Anal. \textbf{262}, 2379, (2012).
\bibitem{acta} P. Garbaczewski, \textit{Dynamics of confined L\'{e}vy flights in terms of (L\'{e}vy) semigroups},
Acta Phys. Pol. {\bf B 43}, 977, (2012)
\bibitem{olkiewicz} P. Garbaczewski and R. Olkiewicz, {\it Feynman-Kac kernels in Markovian
  representations of the Schr\"{o}dinger interpolating dynamics}, J. Math. Phys. {\bf 37}, 732, (1996)
\bibitem{brockmann} D. Brockmann and I. M. Sokolov, \textit{L\'{e}vy flights in external force fields:
From models to equations},  Chemical  Physics,  {\bf 284}, 409, (2002)
\bibitem{SG} P. Garbaczewski and  V. A.  Stephanovich, \textit{L\'{e}vy flights in inhomogeneous environments},
Physica A \textbf{389}, 4419,  (2010).
\bibitem{gar} P. Garbaczewski  and  V. Stephanovich, \textit{L\'{e}vy targeting and the principle
of detailed balance},  Phys. Rev.  E {\bf 84}, 011142 (2011)
\bibitem{risken} H. Risken, {\it The Fokker-Planck Equation}, (Springer-Verlag, Berlin, 1989)
\bibitem{GK} P. Garbaczewski, W. Karwowski, \textit{Impenetrable
barriers and canonical quantization}, Am. J. Phys. \textbf{72}, (2004) 924-933.
\bibitem{BBC} P. Bader, S. Blanes and  F. Casas, \textit{Solving the Schr\"{o}dinger eigenvalue problem by the imaginary
time propagation technique using splitting methods with complex
coefficients}, J. Chemical Physics \textbf{139}, 124117 (2013).
\bibitem{Auer} J. Auer and E. Krotschek, \textit{A fourth-order real-space algorithm for solving local Schr\"{o}dinger equations},
 J. Chem. Phys. {\bf 115},  6841, (2001).
\bibitem{GS} P. Garbaczewski, V. Stephanovich, \textit{L\'{e}vy flights and nonlocal quantum dynamics},
J. Math. Phys. \textbf{54}, (2013) 072103.
\bibitem{ZG} M. ¯aba, P. Garbaczewski, \textit{Solving fractional Schr\"{o}dinger-type spectral problems:
Cauchy oscillator and Cauchy well},
 J. Math. Phys. \textbf{55}, (2014) 092103.
\bibitem{klafter} I. Eliazar and J. Klafter,  \textit{L\'{e}vy-driven Langevin systems:
 Targeted stochasticity},  J. Stat. Phys. {\bf 111}, 739, (2003)
\bibitem{fogedby} S. Jespersen, R. Metzler and H. C. Fogedby, \textit{L\'{e}vy flights in external force
 fields: Langevin and fractional Fokker-Planck equations and their solutions},
 Phys. Rev. E{\bf 59}, 2736, (1999)
\bibitem{B} R. Ba\~{n}uelos and  T. Kulczycki, {\it The Cauchy process and the Steklov problem},
J. Funct. Anal. {\bf 211}, 355, (2004).
\bibitem{B1} R. Ba\~{n}uelos, T. Kulczycki and  P. J. Mendez-Hernandez, {\it On the shape of the
ground state eigenfunction for stable processes},
Potential Analysis, {\bf  24}, 205, (2006).

\end{thebibliography}
\end{document}